\documentstyle[preprint,aps,prd,epsfig]{revtex}

\def\lprox{\mathrel{\raise .3ex\hbox{$<$\kern-
.75em\lower1ex\hbox{$\sim$}}}}

\def\gprox{\mathrel{\raise .3ex\hbox{$>$\kern-
.75em\lower1ex\hbox{$\sim$}}}}

\def\be{\begin{equation}}
\def\ee{\end{equation}}	
\def\ben{\begin{eqnarray}}
\def\een{\end{eqnarray}}

\begin{document}

\draft

\title{\bf Lightcone  fluctuations in flat spacetimes with nontrivial topology}

\author{Hongwei Yu \footnote{e-mail: hwyu@cosmos2.phy.tufts.edu} 
        and L. H. Ford \footnote{e-mail: ford@cosmos2.phy.tufts.edu}}

\address{Institute of Cosmology, Department of Physics and Astronomy\\
	Tufts University, Medford, MA 02155, USA}

\date{\today}

\maketitle

\tightenlines

\begin{abstract}
The quantum lightcone fluctuations in  flat spacetimes with compactified 
spatial dimensions or with boundaries are examined. The discussion is based
upon a model in which the source of the underlying metric fluctuations is taken
 to be quantized linear perturbations of the gravitational field.  General 
expressions are derived, in the transverse trace-free gauge, for the 
summation of  graviton polarization tensors, and for vacuum graviton 
two-point functions.
Because of the fluctuating light cone, the flight time of photons between a
 source and a detector may be either longer or shorter than the light 
propagation time in the background classical spacetime.  We calculate the mean
 deviations from the classical propagation time of photons due to the changes 
in the topology of the flat spacetime.   These deviations are in 
general larger in the directions in which topology changes occur and are
typically of the order of the Planck time,  
but they can get larger as the travel distance increases.

\end{abstract}

\pacs{PACS number(s): 04.60.-m, 04.62.+v}


\section{ Introduction}
 
 The existence of fixed lightcone structures is one of the characteristics of classical gravitational theory.
Lightcones are basically hypersurfaces which distinguish timelike separation from spacelike separation and 
divide spacetime into causally distinct regions. However, if gravity is to be quantized, it is natural to 
expect that the quantum metric fluctuations would smear out the lightcone, 
 and the concept of a fixed lightcone structure has to be abandoned. 
Based upon the observation that the ultraviolet divergences of quantum field theory arise from the 
light cone singularities of two-point functions, and that quantum fluctuations of the spacetime metric ought
 to smear out the light cone, thus possibly removing these singularities, Pauli\cite{Pauli} conjectured many years
 ago that the ultraviolet 
divergences of quantum field theory might be removed if gravity is quantized. This idea was further explored
by several other authors \cite{Deser,DW,ISS}. At present time, this conjecture remains unproven. 
If lightcones fluctuate, so do  horizons,  which are, of course, lightcones. 
The horizon fluctuations could then presumably lead to information leakage across the black hole in a way 
that is not allowed by classical physics.
 Bekenstein and Mukhanov \cite{MB} have suggested that horizon fluctuations could result in discreteness of
 the  spectrum of black holes. Since the existence of black hole  horizons is the origin of the so-called black 
hole information 
paradox, which has been widely discussed in the literature but still remains to be resolved,  the 
study of lightcone fluctuations might help us better understand the problem.

 Recently the problem of light cone fluctuations has been investigated 
\cite{Ford95,Ford96}
in a model of quantum linearized theory of gravity,  where the fluctuations are produced by gravitons propagating on
 a  background spacetime. The lightcone is smeared out if the linearized gravitational perturbations are quantized. 
 It has been demonstrated that
gravitons in a quantum state, such as a squeezed vacuum state,  or a thermal state,  can produce light cone fluctuations, thus smearing out the light cone.  Because of the fluctuating light cone,  
the propagation time of a classical light pulse over distance $r$ is no longer 
precisely $r$, but undergoes fluctuations around a mean value of $r$. The 
fluctuations in the photon arrival time can also be understood as fluctuations
in the velocity of light.    This model
 has  been applied to study the quantum cosmological and black hole horizon fluctuations \cite{Ford97}.  It is interesting to note that recently, the quantum gravitational metric fluctuations have also 
been discussed within a different context, i.e., a Liouville string formulation of quantum 
gravity \cite{AEMN,EMN} . 
 In this paper
we shall examine light cone fluctuations in flat spacetime with nontrivial topology based upon the model proposed in 
Ref\cite{Ford95} . In Sec. II, we review the basic formalism and examine its 
gauge invariance, then derive  general expressions for the vacuum graviton
 two-point functions in the transverse trace-free gauge. In Sec. III we study the light cone fluctuations in
flat spacetimes   with a compactified spatial dimension, and with a single plane boundary. Our results are summarized and
discussed in Sec. VI.

 
\section{ Basic formalism and graviton two-point function in transverse trace-free gauge}


Let us consider a flat background spacetime  with a linearized perturbation 
$h_{\mu\nu}$ propagating upon it , so the spacetime metric
may be written as
\begin{equation}
ds^2 = g_{\mu\nu}dx^\mu dx^\nu = (\eta_{\mu\nu} +h_{\mu\nu})dx^\mu dx^\nu
= dt^2 -d{\bf x}^2 + h_{\mu\nu}dx^\mu dx^\nu \, .  \label{eq:metric}
\end{equation}
Let $\sigma(x,x')$ be one half of the squared geodesic separation
for any pair of spacetime points $x$ and $x'$,   and $\sigma_0(x,x')$ 
be the corresponding quantity in the flat background . We can expand, 
in the presence of the perturbation, $\sigma(x,x')$ in powers of
 $h_{\mu\nu}$ as
\begin{equation}
\sigma = \sigma_0 + \sigma_1 + \sigma_2 + \cdots \, , \label{eq:sigma}
\end{equation}
where $\sigma_1$ is first order in $h_{\mu\nu}$, etc. We now suppose
that the linearized perturbation $h_{\mu\nu}$ is quantized, and that the
quantum state $|\psi \rangle$ is a ``vacuum'' state in the sense that
we can decompose $h_{\mu\nu}$ into positive and negative frequency parts
$h^{+}_{\mu\nu}$ and $h^{-}_{\mu\nu}$, respectively, such that
\begin{equation}
h^{+}_{\mu\nu} |\psi \rangle =0, \qquad \langle \psi|h^{-}_{\mu\nu} =0 \,.
\end{equation}
It follows immediately that
\begin{equation}
\langle  h_{\mu\nu} \rangle =0
\end{equation}
in state $|\psi \rangle$. In general, however, 
$\langle (h_{\mu\nu})^2 \rangle_R \not= 0$, where the expectation value is
understood to be suitably renormalized. This reflects the quantum metric
fluctuations.


\subsection{Basic formalism and gauge invariance}


If we  average the retarded Green's function, $G_{ret}(x,x') $, for a massless scalar field,  over 
quantized metric fluctuations, we get \cite{Ford95} 

\begin{equation}
\Bigl\langle G_{ret}(x,x') \Bigr\rangle = 
{{\theta(t-t')}\over {8\pi^2}} \sqrt{\pi \over {2\langle \sigma_1^2 \rangle}}
\; \exp\biggl(-{{\sigma_0^2}\over {2\langle \sigma_1^2 \rangle}}\biggr)\, .
                                           \label{eq:retav}
\end{equation}
This form is valid for the case in which $\langle \sigma_1^2 \rangle > 0$. 
It reveals that the delta-function behavior of the
classical Green's function, $G_{ret}$, has been smeared out into a Gaussian
function peaked around the classical lightcone. This smearing can be understood as due to the fact
 that photons may be either slowed down or speeded up by the
light cone fluctuations. Photon propagation now becomes a statistical phenomenon, with some photons traveling slower than
the light on the classical spacetime, and others traveling faster.  Note that the Gaussian function 
in Eq.~(\ref{eq:retav}) is symmetrical about the classical light cone,  $ \sigma_0=0$, and so the quantum
fluctuations are equally likely to produce a time advance as a time delay.

Light cone fluctuations are in principle observable. It has been shown, by considering light pulses 
between a source and a detector separated by a distance $r$, that
the mean deviation from the classical propagation time is related to $\langle \sigma_1^2 \rangle$
by \cite{Ford95}
\be
\Delta t= {\sqrt{\langle \sigma_1^2 \rangle}\over r}\,.
\label{eq:MDT}
\ee
Note, however, that $\Delta t$ is the ensemble averaged deviation, not necessarily the expected variation in
flight time of two photons emitted close together in time. The latter can be much smaller than $\Delta t$ due 
to the fact that the gravitational field may not fluctuate significantly in the interval between the two photons.
This point is discussed in detail in Ref.~\cite{Ford96}.
 In order to find $\Delta t$ in a particular situation, 
we need to calculate the quantum expectation value  $\langle \sigma_1^2 \rangle$ in a chosen 
quantum state. For this purpose,  we first have to compute $\sigma_1$ for a given classical perturbation
along a certain geodesic, then average $\sigma_1^2$ over the quantized metric perturbation. If
we consider a null geodesic specified by
\be
dt^2=d{\bf x}^2-h_{\mu\nu}dx^{\mu}dx^{\nu},
\ee
then by following the same steps as those of Ref.\cite{Ford95}, we can show that in a general gauge
\begin{equation}
\sigma_1 = {1\over 2}\Delta r \int_{r_0}^{r_1}
                         h_{\mu\nu} n^{\mu} n^{\nu}\,dr\,,
\end{equation}
and
\begin{equation}
\langle \sigma_1^2 \rangle = {1\over 4}(\Delta r)^2
\int_{r_0}^{r_1} dr \int_{r_0}^{r_1} dr'
\:\, n^{\mu} n^{\nu} n^{\rho} n^{\sigma}
\:\, \langle h_{\mu\nu}(x) h_{\rho\sigma}(x') \rangle_R \,.
\end{equation}
Here $ dr=|d{\bf x}|$,  $\Delta r=r_1-r_0$ and $ n^{\mu} =dx^{\nu}/dr$. 
The  graviton two-point function,
 $\langle h_{\mu\nu}(x) h_{\rho\sigma}(x') \rangle_R$,
is understood to be renormalized, so that it is finite when $x=x'$ and vanishes
when the quantum state of the gravitons is the Minkowski vacuum state.

 A few comments on the derivation of Eq.~(\ref{eq:retav}) are in order here.
It is obtained by averaging the Fourier representation of a $\delta$-function.
It may come as a surprize that although we started with an analytic expansion 
of $\sigma$ in powers of $h_{\mu\nu}$, the result is not analytic as
$\langle \sigma_1^2 \rangle \rightarrow 0$. This arises because we use the 
first order expansion of $\sigma$ in the argument of an exponential function,
but afterwards retain all powers of  $h_{\mu\nu}$. One can reasonably ask
whether this is a valid procedure. A test of the self-consistency is to
retain the $\sigma_2$ term and then follow the same procedure. The result is
Eq.~(\ref{eq:retav}) with $\sigma_0^2$ replaced by $\sigma_0^2 + \sigma_2$.
This has the same physical interpretation as before; the only effect of the 
$\sigma_2$ part is to shift the location of the mean lightcone. Thus in
this order we encounter the backreaction of the gravitons in perturbing the 
original classical geometry to a new classical geometry. Although this is 
less than a complete demonstration of the validity of Eq.~(\ref{eq:retav}),
it does indicate that it arises from a self-consistent calculation. In any
case, the only result that we really need in the remainder of this paper is 
Eq.~(\ref{eq:MDT}), which may be derived either from Eq.~(\ref{eq:retav}),
or else more directly by averaging the square of Eq.~(\ref{eq:sigma}).

Let us now turn to the question of the gauge invariance of the formalism. 
 Under a gauge transformation specified by
\be
x^{\prime \mu}=x^{\mu}+\xi^{\mu}(x),
\ee
\be
h^{\prime }_{\mu\nu}(x')= h_{\mu\nu}(x)-\xi_{(\mu,\nu)}(x)\,,
\ee
where $\xi^{\mu}(x)$ is of  order $h_{\mu\nu}$, the quantities  $\sigma_1$ and
$\langle \sigma_1^2 \rangle$ are not in general invariant. However, we can show that this is due to the fact
that $\Delta t$ is a coordinate time interval rather than a proper time interval.
To better understand the gauge invariance, let us examine a situation in which a light
signal travels between two points in space labeled by $P$ and $Q$,  with a classical metric perturbation 
$h_{\mu\nu}$ in the intervening region, as illustrated in Fig. 1. 
\begin{figure}[hbtp]
\begin{center}
\leavevmode\epsfxsize=1.6in\epsfbox{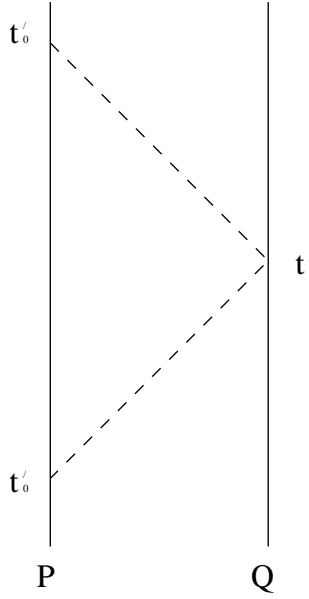}
\end{center}
\caption{ A light ray ( dashed line ) makes a round trip travel between two points, P and Q, in space.}
\label{fig=fig1}
\end{figure}
 For simplicity, let us assume
that the propagation is in the $x$-direction. We shall look at the travel time in two different gauges,
 or coordinate systems, primed and unprimed. 
For  light rays traveling in $x$ direction, we have
\ben
{dt\over dx }& =& \pm \sqrt{ 1-h_{\mu\nu}(x){dx^{\mu}\over dx}{dx^{\nu}\over dx}}\,\nonumber\\
&&\approx \pm 1 \mp{1\over2}h_{\mu\nu}(x){dx^{\mu}\over dx}{dx^{\nu}\over dx}\,.
\een
Here the upper sign is used for outgoing light rays and the lower sign for incoming rays. 
So,  one way travel time $\delta t$ in the unprimed gauge is 
\ben
\delta t_{P\rightarrow Q}&=&\int^{x_Q}_{x_P}\,dx\,-{1\over 2}\int^{x_Q}_{x_P}\,h_{\mu\nu}(x){dx^{\mu}\over dx}
{dx^{\nu}\over dx}\,dx\nonumber\\
&& =\int^{x_Q}_{x_P}\,dx\,-{1\over 2}\int^{x_Q}_{x_P}\,h^{'}_{\mu\nu}(x'){dx^{\mu}\over dx}
{dx^{\nu}\over dx}\,dx -{1\over 2}\int^{x_Q}_{x_P}\,\xi_{(\mu,\nu)}(x){dx^{\mu}\over dx}
{dx^{\nu}\over dx}\,dx\,, \nonumber\\
\een
which, within  the linearized theory, can approximated as
\ben
\delta t_{P\rightarrow Q}&=&
\int^{x_Q}_{x_P}\,dx\,-{1\over 2}\int^{x^{'}_Q}_{x^{'}_P}\,h^{'}_{\mu\nu}(x'){dx^{'\mu}\over dx'}
{dx^{'\nu}\over dx'}\,dx' -{1\over 2}\int^{x_Q}_{x_P}\,\xi_{(\mu,\nu)}(x){dx^{\mu}\over dx}
{dx^{\nu}\over dx}\,dx, \nonumber\\
&&=\int^{x_Q}_{x_P}\,dx\,-{1\over 2}\int^{x^{'}_Q}_{x^{'}_P}\,h^{'}_{\mu\nu}(x'){dx^{'\mu}\over dx'}
{dx^{'\nu}\over dx'}\,dx' -\int^{x_Q}_{x_P}\,{d\xi_x\over dx}\,dx-\int^{x_Q}_{x_P}\,{d\xi_t\over dx}\,dx
\nonumber\\
&&=x_Q(t')-\xi_x(Q,t')-(\,x_P(t_0)-\xi_x(P,t_0)\,)-\xi_t(Q,t')+\xi_t(P,t_0)\nonumber\\
&&\quad -{1\over 2}\int^{x^{'}_Q}_{x^{'}_P}\,h^{'}_{\mu\nu}(x'){dx^{'\mu}\over dx'}{dx^{'\nu}\over dx'}\,dx'\,,\nonumber\\  
\een
where we have used the fact $ dt/dx=1$ for outgoing light rays within our approximation. 
Similarly, we have
\ben
\delta t_{Q\rightarrow P}&=&-\int^{x_P}_{x_Q}\,dx\,+{1\over 2}\int^{x_P}_{x_Q}\,h_{\mu\nu}(x){dx^{\mu}\over dx}
{dx^{\nu}\over dx}\,dx\nonumber\\
&& =\int^{x_Q}_{x_P}\,dx\,-{1\over 2}\int^{x_Q}_{x_P}\,h^{'}_{\mu\nu}(x'){dx^{\mu}\over dx}
{dx^{\nu}\over dx}\,dx +{1\over 2}\int^{x_P}_{x_Q}\,\xi_{(\mu,\nu)}(x){dx^{\mu}\over dx}
{dx^{\nu}\over dx}\,dx, \nonumber\\
&&=\int^{x_Q}_{x_P}\,dx\,-{1\over 2}\int^{x^{'}_Q}_{x^{'}_P}\,h^{'}_{\mu\nu}(x'){dx^{'\mu}\over dx'}
{dx^{'\nu}\over dx'}\,dx' +\int^{x_P}_{x_Q}\,{d\xi_x\over dx}\,dx-\int^{x_P}_{x_Q}\,{d\xi_t\over dx}\,dx
\nonumber\\
&&=x_Q(t')-\xi_x(Q,t')-(\,x_P(t_0')-\xi_x(P,t_0')\,)+\xi_t(Q,t')-\xi_t(P,t_0')\nonumber\\
&&\quad -{1\over 2}\int^{x^{'}_Q}_{x^{'}_P}\,h^{'}_{\mu\nu}(x'){dx^{'\mu}\over dx'}{dx^{'\nu}\over dx'}\,dx'\,,\nonumber\\  
\een
 using the fact that for incoming light rays, $dt/dx=-1$. 
 Note that 
\be
x'_Q(t)=x_Q(t)-\xi_x(Q,t)\,,
\ee
\be
x'_P(t)=x_P(t)-\xi_x(P,t)\,,
\ee
 so
\be
\delta t_{P\rightarrow Q} =\delta t'_{P\rightarrow Q}-\xi_t(Q,t')+\xi_t(P,t_0)\,,
\ee
and 
\be
\delta t_{Q\rightarrow P}=\delta t'_{Q\rightarrow P}+\xi_t(Q,t')-\xi_t(P,t_0')\,.
\ee
It follows that  the round trip travel time is 
\be
\Delta t=\delta t_{P\rightarrow Q}+\delta t_{Q\rightarrow P}=\Delta t' +\xi_t(P,t_0)-\xi_t(P,t_0')\,.
\label{eq:corrditime}
\ee

Therefore, the one way travel times, $\delta t_{P\rightarrow Q}$ and 
$\delta t_{Q\rightarrow P} $ are, in general, not invariant unless both the source and the  
detector are outside the regions where gravitational perturbations $h_{\mu\nu}$ are non-zero. 
In that case, it is physically 
reasonable to set $\xi(P,t)$ and $\xi(Q,t)$ to zero. Similarly,  the round trip time $\Delta t$ 
 is invariant only if the source ( it also acts as a detector in this case ) is outside of the 
gravitational perturbations.

 However, it is interesting to note that the round trip proper time interval for the source,  
$\Delta \tau$, is gauge invariant. Denote the proper
 time intervals in two different gauges by $\Delta \tau$ and $\Delta \tau'$, and keep in mind the fact that on the
 world line of the source, generally, ${dx^i\over dt}<<1$.  We then have
\ben
\Delta \tau'&=&\int \sqrt{1+h'_{00}}\,dt'=\int dt'+{1\over 2}\int h'_{00} dt'\nonumber\\
&&=\Delta t'+{1\over 2}\int h_{00}dt-\int {d \xi_t\over dt}\, dt\nonumber\\
&&=\Delta t'+\xi_t(P,t_0)-\xi_t(P,t_0')
+{1\over 2}\int h_{00} dt \nonumber\\
&&=\Delta t+{1\over 2}\int h_{00} dt=\Delta \tau\,, \nonumber\\
\een
where we have used Eq.~(\ref{eq:corrditime}).
This shows that we should really consider how proper time rather than the  coordinate time is
affected by light cone fluctuations. 
However, the calculation of the  proper time in a 
general gauge is a rather difficult task, because the source (and detector) may not be at rest with 
respect to the chosen coordinate system, and thus  in general the emission and the subsequent reception 
may not 
happen at the same point in space. To find the proper time, we have to integrate along the 
geodesic between two events, the emission and the subsequent reception.  In general, there is  a 
Doppler shift due to fluctuations in the positions of the source and the mirror.   However, the analysis 
can be greatly simplified if we adopt the transverse-tracefree ( TT ) gauge, which is  
 specified by the conditions
\begin{equation}
h^j_j = \partial_j h^{ij} = h^{0\nu} = 0\, . \label{eq:the TT}
\end{equation}
To see this, let us examine the  geodesic equations for a test particle
\be
{d^2x^{\mu}\over d^2\lambda}=\,-\Gamma^{\mu}_{\rho\sigma}\,{dx^{\rho}\over d\lambda}
{dx^{\rho}\over d\lambda}\,,
 \ee
which, when written in term of derivatives with respect to coordinate time $t$, becomes
\be
{d^2x^{\mu}\over d^2 t}\,+\Gamma^{\mu}_{\rho\sigma}\,{dx^{\rho}\over dt}{dx^{\sigma}\over dt}\,
-\Gamma^{t}_{\rho\sigma}\,{dx^{\mu}\over dt}\,{dx^{\rho}\over dt}{dx^{\sigma}\over dt}=0\,.
 \ee
For a non-relativistic test particle,   ${dx^i\over dt}<<1$,  so,  to the leading order, 
\be
 {d^2x^i\over d^2 t}\,\approx \Gamma^{i}_{tt}\,.
 \ee

 But, in the TT gauge,  $\Gamma^{i}_{tt}=0$. Therefore, from the above equation, 
 we can see that if the test particle is at rest at $t=0$, then it will subsequently always remain at rest
 \cite{MTW}.   So, if we are considering the 
emission and reflection of a light signal between two points (particles) in the TT gauge, 
then the proper time $\delta \tau$ between emission and reception (after 
reflection) of the signal is related with the coordinate time by
\be
\delta \tau=\int \sqrt{g_{tt}}dt=\int \sqrt{(1+h_{00})}dt=\int dt=\delta t\,.
\ee
Here we have appealed to the fact that $h_{00}=0$ in the TT gauge. Therefore, the coordinate time for the round trip in the TT
 gauge is the proper time, and $\Delta t $ calculated from Eq.~(\ref{eq:MDT}) in the TT guage is actually a
 gauge invariant quantity.
In this gauge,  the mean squared fluctuation in the geodesic 
interval function reduces to 
\begin{eqnarray}
\langle \sigma_1^2 \rangle &&= {1\over 4}(\Delta r)^2
\int_{r_0}^{r_1} dr \int_{r_0}^{r_1} dr'
\:\, n^i n^j n^k n^m
\:\, \langle h_{ij}(x) h_{km}(x') \rangle_R \, \nonumber\\
&&= {1\over 8}(\Delta r)^2
\int_{r_0}^{r_1} dr \int_{r_0}^{r_1} dr'
\:\, n^i n^j n^k n^m
\:\, \langle h_{ij}(x) h_{km}(x')+ h_{ij}(x') h_{km}(x) \rangle_R \,.
\label{eq: interval}
\end{eqnarray}
Here $ n^i =dx^i/dr$ is the unit three-vector defining the spatial 
direction of the geodesic.

\subsection{Graviton two-point function in transverse trace-free gauge}

If we work in the TT gauge, the gravitational perturbations have only spatial components  $h_{ij}$ 
and they may be quantized using a plane wave expansion
 as
\begin{equation}
h_{ij} = \sum_{{\bf k},\lambda}\, [a_{{\bf k}, \lambda} e_{ij} ({{\bf k}, \lambda})
  f_{\bf k} + H.c. ].
\end{equation}
Here H.c. denotes the Hermitian conjugate, $\lambda$ labels the polarization
states, and
\be
f_{\bf k} = (2\omega (2\pi)^3)^{-{1\over 2}}  e^{i({\bf k \cdot x} -\omega t)}
\ee
is the mode function, where
\be
\omega=|{\bf k}|, \qquad |{\bf k}|=(k_x^2+k_y^2+k_z^2)^{{1\over2}},
\ee
 and the $e_{\mu\nu} ({{\bf k}, \lambda})$
are polarization tensors. ( Units in which $32\pi G =1$, where $G$ is
Newton's constant and in which $\hbar =c =1$  will be used in this paper.)

Now we shall first calculate the Minkowski spacetime Hadamard function for gravitons in the
 transverse tracefree gauge. Let us define

\begin{equation}
G^{(1)}_{ijkl}(x,x')=\langle 0| h_{ij}(x) h_{kl}(x')+ h_{ij}(x') h_{kl}(x)|0 \rangle \,.
\end{equation}
Then we have 
\begin{equation}
 G^{(1)}_{ijkl}(x,x')=\frac{2 Re}{(2\pi)^3}\int\,d^3{\bf k}\sum_{\lambda}
\, e_{ij} ({{\bf k}, \lambda}) e_{kl} ({{\bf k}, \lambda})
{1\over{2 \omega}}e^{i{\bf k} \cdot({\bf x}-{\bf x'})}e^{-i\omega(t-t')}\,.
\end{equation}
 Equation~(\ref{eq:polsum}) in the Appendix for the summation of polarization tensors in the 
transverse tracefree gauge gives

\begin{eqnarray}
\sum_{\lambda}\, e_{ij} ({{\bf k}, \lambda}) e_{kl} ({{\bf k}, \lambda})&=&\delta_{ik}\delta_{jl}
+\delta_{il}\delta_{jk}-\delta_{ij}\delta_{kl}
+\hat k_i\hat k_j \hat k_k\hat k_l+\hat k_i \hat k_j \delta_{kl} \nonumber\\
&&+\hat k_k \hat k_l \delta_{ij}-\hat k_i \hat k_l \delta_{jk}
-\hat k_i \hat k_k \delta_{jl}-\hat k_j \hat k_l \delta_{ik}-\hat k_j \hat k_k \delta_{il}\,,
\end{eqnarray}
where 
\begin{equation}
\hat k_i=\frac{ k_i}{ k}\,.
\end{equation}
 We find that $  G^{(1)}_{ijkl}(x,x')$ can be expressed as \cite{Footnote}

\begin{eqnarray}
 G^{(1)}_{ijkl}(x,x')&&=2 Re \,(\delta_{ik}\delta_{jl}+\delta_{il}\delta_{jk}-\delta_{ij}\delta_{kl} + D_{ij} )
\times {1\over{(2\pi)^3}}\int\, d^3{\bf k}{1\over{2 \omega}}
e^{i{\bf k} \cdot({\bf x}-{\bf x'})}e^{-i\omega(t-t')}\nonumber\\
&&=2 Re \,(\delta_{ik}\delta_{jl}+
\delta_{il}\delta_{jk}-\delta_{ij}\delta_{kl} +D_{ij})\times \langle 0|\phi(x)\phi(x') |0 \rangle\,,\
\end{eqnarray}
where we have defined a formal operator 
\begin{equation}
D_{ij}=\left( {\partial_i\partial^{\prime}_j\over{\nabla^2}}\delta_{kl}+ 
{\partial_k\partial^{\prime}_l\over{\nabla^2}}\delta_{ij}
- {\partial_i\partial^{\prime}_k\over{\nabla^2}}\delta_{jl}
-  {\partial_i\partial^{\prime}_l\over{\nabla^2}}\delta_{jk}
- {\partial_j\partial^{\prime}_l\over{\nabla^2}}\delta_{ik}
-{\partial_j\partial^{\prime}_k\over{\nabla^2}}\delta_{il} 
+{\partial_i\partial_j^{\prime}\partial_k\partial_l^{\prime}\over\nabla^4}
 \right),
\end{equation} 
and $ \langle 0|\phi(x)\phi(x') |0 \rangle$ is the usual scalar field two-point function. Here the formal 
operator $\nabla^{-2}$ should be understood in the sense of a Green's function, but when we do our 
calculations in  momentum space its effect is to bring in a factor of $k^{-2}$.

The combination of these results with Eq.~(\ref{eq: interval} ) gives

\begin{equation}
 \langle \sigma_1^2 \rangle = { 1\over 4}(\Delta r)^2
\int_{r_0}^{r_1} dr \int_{r_0}^{r_1} dr'
\:\,\left( 1-{ 2 ({\bf \nabla }\cdot {\bf n})({\bf \nabla}^{\prime} \cdot {\bf n})\over{\nabla^2} }
+{  ({\bf \nabla }\cdot {\bf n})^2({\bf \nabla}^{\prime} \cdot {\bf n})^2\over{\nabla^4} }
 \right) \langle \phi(x)\phi(x')  \rangle_R\,.\nonumber\\
\end{equation}
 Introduce two functions $F_{ij}(x,x')$ and $H_{ijkl}(x,x')$ by

\begin{eqnarray}
 	F_{ij}(x,x')&&= Re  {\partial_i\partial^{\prime}_j\over{\nabla^2}} 
\langle 0|\phi(x)\phi(x') |0 \rangle \nonumber\\
&&= Re  {\partial_i\partial^{\prime}_j\over{\nabla^2}}
{1\over{(2\pi)^3}}\int\, d^3{\bf k}{1\over{2 \omega}}e^{i{\bf k} \cdot({\bf x}-{\bf x'})}e^{-i\omega(t-t')}\nonumber\\
&&={Re\over{(2\pi)^3}}\int\, d^3{\bf k}{k_ik_j\over{2 \omega^3}}e^{i{\bf k} \cdot({\bf x}-{\bf x'})}e^{-i\omega(t-t')} \,,
\end{eqnarray}
and
\begin{eqnarray}
 	H_{ijkl}(x,x')&&= Re  {\partial_i\partial^{\prime}_j\partial_k\partial_{l}^{\prime}
\over{\nabla^4}} \langle 0|\phi(x)\phi(x') |0 \rangle \nonumber\\
&&= Re  {\partial_i\partial^{\prime}_j\partial_k\partial_l^{\prime}\over{\nabla^4}}
{1\over{(2\pi)^3}}\int\, d^3{\bf k}{1\over{2 \omega}}e^{i{\bf k} \cdot({\bf x}-{\bf x'})}e^{-i\omega(t-t')}\nonumber\\
&&={Re\over{(2\pi)^3}}\int\, d^3{\bf k}{k_i k_j k_k k_l\over{2 \omega^5}}e^{i{\bf k} \cdot({\bf x}-{\bf x'})}e^{-i\omega(t-t')} \,.
\end{eqnarray}
 $G^{(1)}_{ijkl}$ can be expressed as
\ben
G^{(1)}_{ijkl}&=&
2F_{ij}\delta_{kl} +2F_{kl}\delta_{ij}-2F_{ik}\delta_{jl}-2F_{il}\delta_{jk}-2F_{jl}\delta_{ik}
-2F_{jk}\delta_{il}+2H_{ijkl}\nonumber\\
&&\quad +2D^{(1)}(x,x')(\delta_{ik}\delta_{jl}+\delta_{il}\delta_{jk}-\delta_{ij}\delta_{kl}),
\een
where
\be
D^{(1)}(x,x')= -{1\over{8\pi^2\sigma_0^2}}
\ee
is the usual Hadamard function for massless scalar fields with  
$2\sigma_0^2=(t-t')^2-({\bf x-x' })^2$, and  $F_{ij}(x,x')$ and $H_{ijkl}(x,x')$, which will be
 calculated in the Appendix, are given by
\be
F_{ij}(x,x')=-{1\over{(2\pi)^2}}\partial_i\partial_j'\,\left[{1\over 2}\ln ( R^2-\Delta t^2 )
+{\Delta t \over 4R}\ln \Bigg( \frac{R+\Delta t}{R-\Delta t}\Bigg)^2  \right]\,,
\label{eq:Ffunc}
\ee
and
\ben
H_{ijkl}(x,x')&=&{1\over 96\pi^2}\partial_i\partial_j'\partial_k\partial_l'\,
\Bigg[ (R^2+3\Delta t^2)\ln (R^2-\Delta t^2)^2\nonumber\\
\quad && + \left(3R\Delta t+{\Delta t^3\over R}\right)\ln \Bigg( \frac{R+\Delta t}{R-\Delta t}\Bigg)^2\Bigg]\nonumber\,.\\
\label{eq:Hfunc}
\een
Here $R=|{\bf x}-{\bf x}'|$. 


\section{Lightcone fluctuations in flat spacetime with nontrivial topologies or boundaries}

In this section, we study lightcone fluctuations in two cases: 
  flat spacetime with a compactified spatial section,  and with a single plane boundary.


\subsection{ Flat spacetime with a compactified spatial section}
 
  Let us now assume that the spacetime is flat but compactified in the $z$ direction with a 
periodicity length
$L$ ( ``circumference of the universe'' ). This means the spatial points $z$ and $z+L$ are identified.
 The effect of the space closure is to restrict the field modes to a discrete set
\be
f_{\bf k} = (2\omega (2\pi)^2L)^{-{1\over 2}}  e^{i({\bf k \cdot x} -\omega t)}
	\label{eq:mode1}
\ee
with 
\be
k_z={2\pi n\over L}, \qquad n=0,\pm 1, \pm 2, \pm 3,...
\ee
We now analyze the lightcone fluctuations, assuming that the gravitons are in the new vacuum state
$|0_L\rangle$ associated the discrete modes of Eq.~(\ref{eq:mode1} ).

 First consider a light ray  along $z$ direction, i.e. along the direction of compactification 
(Fig.2 ),
propagating from point $(0,0,a)$ to point $(0,0,b)$ in space, 
then, we have from Eq.~(\ref{eq: interval})

\ben
\langle \sigma_1^2 \rangle &&= {1\over 8}(b-a)^2
\int_{a}^{b} dz \int_{a}^{b} dz'
\:\, \langle 0_L| h_{zz}(x) h_{zz}(x')+ h_{zz}(x') h_{zz}(x)|0_L \rangle_R \,\nonumber\\
 &&= {1\over 8}(b-a)^2
\int_{a}^{b} dz \int_{a}^{b} dz'\, G_{zzzz}^{(1) R}(t,0,0,z,\,t',0,0,z').
 	\label{eq: interval1}
\een
Here we have defined 
\ben
&&G^{(1)R}_{zzzz}(x,x')=\langle 0_L| h_{zz}(x) h_{zz}(x')+ h_{zz}(x') h_{zz}(x)|0_L \rangle_R\nonumber\\
&&\qquad=\langle 0_L| h_{zz}(x) h_{zz}(x')+ h_{zz}(x') h_{zz}(x)|0_L \rangle-
\langle 0| h_{zz}(x) h_{zz}(x')+ h_{zz}(x') h_{zz}(x)|0 \rangle\,,
\een
and the integral is to be carried out along the geodesic.
\vskip.25in

\begin{figure}[hbtp]
\begin{center}
\leavevmode\epsfxsize=1.8in\epsfbox{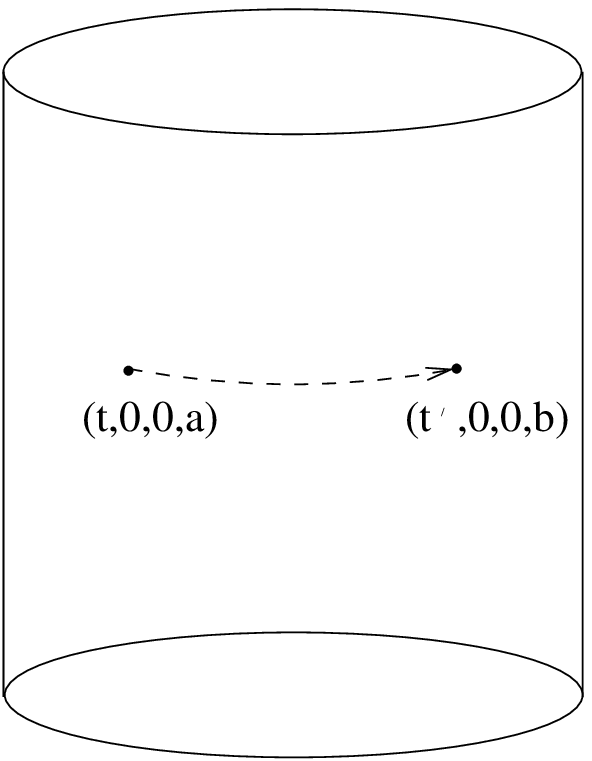}
\end{center}
\caption{ A light ray ( dashed line ) propagates in the direction of compactification in a cylindrical ``universe''
from point $(t,0,0,a)$ to point $ (t',0,0,b)$. Here only two spatial dimensions are plotted. }
\label{fig=fig2}
\end{figure}
If we adopt the notation 
\be
(t,0,0,z,\,t',0,0,z')\equiv(t,z,\,t',z')\,,
\ee
the renormalized two-point function can be found by using the method of images
to be
\ben
 G_{zzzz}^{(1) R}(t,z,\,t',z')&&={\sum_{n=-\infty}^{+\infty}}^{\prime}G_{zzzz}^{(1) }(t,z,\,t',z'+nL)\nonumber\\
&&=2\,{\sum_{n=-\infty}^{+\infty}}^{\prime}\,\Biggl(D^{(1)}(t,z,\, t', z'+nL)-2F_{zz}(t,z,\,t',z'+nL)
\nonumber\\
&&\quad +H_{zzzz}(t,z,\,t',z'+nL)\Biggr)\,,\nonumber\\
\een
where the prime on the summation indicates that the $n=0$ term is omitted. Substituting $R_t=0$
into Eq.~(\ref{eq:tpFunc}) in the Appendix and replacing $\Delta x$ by $\Delta z$,  we have
\ben
 G_{zzzz}^{(1) R}(t,z,\,t',z')&&=-{2\over \pi^2}{\sum_{n=-\infty}^{+\infty}}^{\prime} \Bigg[ \frac{\Delta t^2}{(\Delta z-nL)^4}
+\frac{\Delta t^3}{4(\Delta z-nL)^5}\ln\left( \frac{\Delta z-nL-\Delta t}{\Delta z-nL+\Delta t} \right)^2 
\nonumber\\
&&  - {2\over 3(\Delta z-nL)^2}-\frac{\Delta t}{4(\Delta z-nL)^3}\ln\left( \frac{\Delta z-nL-\Delta t}{\Delta z-nL+\Delta t} \right)^2
\Bigg].
\label{eq:TPF1}
\een

 For the null geodesic
\be
\Delta t=\Delta z,
\ee
 we get, after an evaluation of the integral,
\ben
&&\int_{a}^{b} dz \int_{a}^{b} dz'\, G_{zzzz}^{(1) R}(t,z,\,t',z')|_{\Delta t=\Delta z }\nonumber\\
&&\quad ={1\over12\pi^2 }{\sum_{n=-\infty}^{+\infty}}^{\prime}
\Bigg[{8\epsilon^2 (n^2-2\epsilon^2)\over {(n^2-\epsilon^2)^2}}
 +\frac{(n+2\epsilon)^3}{2(n+\epsilon)^3}\ln\left(1+{2\epsilon\over n}\right)^2
+ \frac{(n-2\epsilon)^3}{2(n-\epsilon)^2}\ln\left(1-{2\epsilon\over n}\right)^2 \Bigg]\nonumber\\
&&\quad ={1\over12\pi^2 }{\sum_{n=1}^{+\infty}}
\Bigg[{16\epsilon^2 (n^2-2\epsilon^2)\over {(n^2-\epsilon^2)^2}}
 +\frac{(n+2\epsilon)^3}{(n+\epsilon)^3}\ln\left(1+{2\epsilon\over n}\right)^2
+ \frac{(n-2\epsilon)^3}{(n-\epsilon)^3}\ln\left(1-{2\epsilon\over n}\right)^2 \Bigg]\nonumber\\
&&\quad\equiv{1\over12\pi^2 }{\sum_{n=1}^{+\infty}}f(n,\epsilon)\nonumber\,,\\
	\label{eq:series1}
\een
where we have defined

\be
\epsilon\equiv{(b-a)\over L}={r\over L}\,,
\ee
and 
\be
 f(n,\epsilon) \equiv {16\epsilon^2 (n^2-2\epsilon^2)\over {(n^2-\epsilon^2)^2}}+
 \frac{(n+2\epsilon)^3}{(n+\epsilon)^3}\ln\left(1+{2\epsilon\over n}\right)^2
+ \frac{(n-2\epsilon)^3}{(n-\epsilon)^3}\ln\left(1-{2\epsilon\over n}\right)^2\,.
\ee

It appears that there is a singularity in the summand $f(n,\epsilon)$ whenever $n=\epsilon$, i.e., 
whenever the distance $r$ is an integer multiple of $L$.  
However this singularity is illusionary, as it should be from a physical point of view since there is nothing special
when $n=\epsilon$. This can be seen if we  expand the summand  at the point $\epsilon=n$ to get
\be
f(n,\epsilon)\approx {19\over 3}+{27\over 4}\ln(3)+\frac{27\ln(3)+68}{8n}(\epsilon-n)+O((\epsilon-n)^2)\,. 
\ee
So, $f(n,\epsilon)$ is finite as $\epsilon$ approaches $n$. 
Note also that $2\epsilon=n$ is also not a singularity. 
The  summation  converges, as the asymptotic form of $f(n.\epsilon)$ as $n\rightarrow \infty$ is 
\be
f(n,\epsilon)\sim {32\epsilon^2\over n^2} + O(n^{-4})\,.
\ee
  However, a generic closed form result for the  summation is  
 hard to find. So we now discuss two special cases. The first
is the one in which
the distance traversed by the light ray is much less than the periodicity length, $ b-a \ll L$.
Then we get

\be
\int_{a}^{b} dz \int_{a}^{b} dz'\, G_{zzzz}^{(1) R}(x,x')
\approx\sum_{n=1}^{+\infty}{8\epsilon^2\over3 \pi^2}{1\over n^2}
={4\epsilon^2\over9}\,.
\ee 
 Substitution of this result into Eq.~(\ref{eq: interval1}) yields
\be
\langle \sigma_1^2 \rangle \approx {r^4\over 18L^2 }.
\ee
Therefore the mean deviation from the classical propagation time is
\be
\Delta t={\sqrt{\langle \sigma_1^2 \rangle }\over r} \approx {1\over 3\sqrt{2} }
\,{r\over L}\,.
\label{eq:t2}
\ee

Since we are working in Natural Units, this result reveals that the mean deviation in travel time
 is less than the
Planck time and grows linearly with increasing $r$ when 
$r$ is  small compared to the periodicity length $L$ of the universe.

If $\epsilon\gg 1$, i.e., $r\gg L$,  the light loops around the `` universe '',  and summation 
Eq~(\ref{eq:series1}) can be approximated by the following integral

\ben
\int_{a}^{b} dz \int_{a}^{b} dz'\, G_{zzzz}^{(1) R}(t,z,\,t',z')&&\approx
{\epsilon\over 12\pi^2}\int_{1/\epsilon}^{\infty}\,dx
\Bigg[\frac{(x+2)^3}{(x+1)^3}\ln\left(1+{2\over x}\right)^2
+ \frac{(x-2)^3}{(x-1)^3}\ln\left(1-{2\over x}\right)^2 \nonumber\\
&& \qquad \qquad \qquad \quad +{16(x^2-2)\over {(x^2-1)^2}}\Bigg]\,.
\een
Evaluating the integral with the aid of the computer algebra package Maple, 
series expanding the result and keeping the leading terms only, we arrive at
\be
\int_{a}^{b} dz \int_{a}^{b} dz'\, G_{zzzz}^{(1) R}(t,z,\,t',z')
 \approx \epsilon-{8\ln(2\epsilon)\over 3\pi^2 }\,.
\ee
Therefore the mean deviation from the classical propagation time is
\be
\Delta t={\sqrt{\langle \sigma_1^2 \rangle }\over r} \approx {1\over 2\sqrt{2}} \,\sqrt{{r\over L}},
\ee
where $r$ is assumed to be much greater than $L$. So  the lightcone fluctuations can, in principle, 
 get as large as one would like if the light ray travels around and around. This is interesting in the sense that
it suggests that a fluctuation which is much greater than the Planck scale could be achieved.

Now we turn to the case where the light ray moves along the direction perpendicular to that of compactification,
for instance, along $x$ direction.
If the light ray travels from point $(a,0,0)$ to point $(b,0,0)$,  as illustrated in Fig. 3,  then

\ben
\langle \sigma_1^2 \rangle &&= {1\over 8}(b-a)^2
\int_{a}^{b} dx \int_{a}^{b} dx'
\:\, \langle 0_L| h_{xx}(x) h_{xx}(x')+ h_{xx}(x') h_{xx}(x)|0_L \rangle_R \,\nonumber\\
 &&={1\over 8}(b-a)^2
\int_{a}^{b} dx \int_{a}^{b} dx'\, G_{xxxx}^{(1)R }(t,x,0,0,\,t',x',0,0), \,\nonumber\\
&&= {1\over 8}(b-a)^2
\int_{a}^{b} dx \int_{a}^{b} dx'\,{\sum_{n=-\infty}^{+\infty}}^{\prime} G_{xxxx}^{(1) }(t,x,0,0,\,t',x',0,nL)\,.
\een

\vskip.15in
\begin{figure}
\begin{center}
\leavevmode\epsfxsize=1.8in\epsfbox{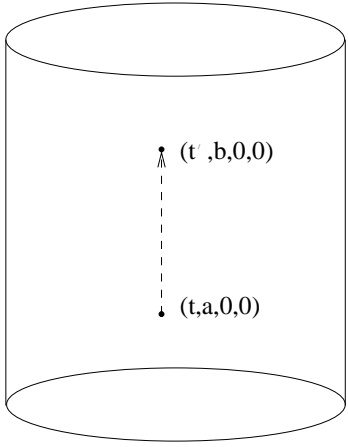}
\end{center}
\caption{  A light ray ( dashed line ) propagates perpendicular to the direction of compactification
 in a cylindrical ``universe''
from point $(t,a,0,0)$ to point $ (t',b,0,0)$. Here only two spatial dimensions are plotted. }
\label{fig=fig3}
\end{figure}
Let us now define 
\be
\rho=x-x',\quad\quad b-a=r,
\ee
then if we use Eq.~(\ref{eq:tpFunc}) in the Appendix and bear in mind the fact that for the light ray $\Delta t=\Delta x$,
we have
\be
 G_{xxxx}^{(1)R }(t,x,0,0,\,t',x',0,nL)\equiv g_1(\rho)+g_2(\rho)\,,
\label{eq:G1}
\ee
 where
\be
g_1 = 2 {\sum_{n=-\infty}^{+\infty}}^{\prime}- {\displaystyle \frac {1}{8\pi^2}} \,{\displaystyle 
\frac {\rho^{2}\,(nL)^{4}}{(\rho^{2} + (nL)^{2})^{4}}} - {1\over3\pi^2 } {\displaystyle \frac {\rho^{6}}{(\rho^{2} + (nL)^{2
})^{4}}} + {\displaystyle \frac {47}{12\pi^2}} \,{\displaystyle 
\frac {\rho^{4}\,(nL)^{2}}{(\rho^{2} + (nL)^{2})^{4}}} \,,
 \ee
 and
\begin{eqnarray}
g_2=-2{\sum_{n=-\infty}^{+\infty}}^{\prime}  \frac {1}{2\pi^2} \,\ln\left( \frac {\sqrt{\rho^{2} + (nL)^2} + \rho}{\sqrt{\rho^{2} + (nL)^{2}} - \rho}\right )^2
&&\Biggl[ -  \frac {1}{16} \, \frac { \rho\,(nL)^6 }{ (\rho^{2} + (nL)^{2})^{(9/2)} }  - \frac {3}{4} \,
 \frac { \rho^3\,(nL)^4 }{ (\rho^{2} + (nL)^{2})^{(9/2)} } \nonumber\\
&&  \quad + \frac {3}{2} \,\frac { \rho^{5}\,(nL)^2 } { (\rho^2 + (nL)^2 )^{(9/2)} }\Biggr]\,.
\nonumber\\ 
\label{eq:G3}
 \end{eqnarray}
We can clearly see that $G_{xxxx}^{(1)R}$ is an even function of $\rho$, so,
\be
\int_{a}^{b} dx \int_{a}^{b} dx'\, G_{xxxx}^{(1)R }(t,x,0,0,\,t',x',0,0)=
2\int_0^r d\rho (r-\rho)(g_1+g_2) \,.
\ee
Performing the integration (integrate by parts for those terms involving logarithmic function), we arrive at
\ben
&&2\int_0^r d\rho (r-\rho)(g_1+g_2)\nonumber\\
&&\quad ={2\over\pi^2 }{\sum_{n=-\infty}^{+\infty}}^{\prime}\Bigg[
-\frac{\epsilon^4}{2(\epsilon^2+n^2)^2}-\frac{\epsilon^2n^2}{4(\epsilon^2+n^2)^2}
\nonumber\\
&&\quad\quad \quad\quad+\frac{8\epsilon^5+
8n^2\epsilon^3+3n^4\epsilon}{24(n^2+\epsilon^2)^{5/2}}
\ln\left({\sqrt{n^2+\epsilon^2}+ \epsilon}\over{\sqrt{n^2+\epsilon^2}- \epsilon} \right)\Bigg]\,,\nonumber\\
	\label{eq: series2}
\een
where $\epsilon=r/L$ as before.
The above series can be shown to be  convergent. 
 Yet a result in closed form is not easy to find.   Let us  first
 examine the case in which $ r\ll L$, where
\be
\int_{a}^{b} dx \int_{a}^{b} dx'\, G_{xxxx}^{(1) R}(x,x')\approx
\sum_{n=1}^{+\infty}{64\epsilon^6\over 45\pi^2}{1\over n^6}
={64\pi^4\epsilon^6\over 45^2\times21}\,.
\ee
Here we have used
\be
 \sum_{n=1}^{+\infty}{1\over n^6}={\pi^6\over 45\times 21}\\.
\ee
Thus the mean deviation from the classical propagation time is 
\be
\Delta t={\sqrt{\langle \sigma_1^2 \rangle }\over r}\approx \sqrt{{{2\over21 }}} {2\pi^2\over 45}\,
\left({r\over L}\right)^3\,.
\ee
This result holds in the small $\epsilon$ regime. 
  The time deviation is much smaller than that for light rays propagating along the compactification direction ( compare with Eq.~(\ref{eq:t2}) ).
 This reveals that light cone fluctuations due to topology change are more likely to be felt in the direction
 of 
compactification than in the transverse direction, if we  perform local 
experiments in which $r$, the distance between the source and the 
detector, is very small as compared to $L$, the periodicity length.

We now turn our attention to the case in which $r\gg L$, i.e., $\epsilon\gg 1$. 
Here it is easy to see that the summation in Eq.~(\ref{eq: series2}) can be approximated by the
following integral 

\ben
&&\int_{a}^{b} dx \int_{a}^{b} dx'\, G_{xxxx}^{(1) R}(t,x,0,0,\,t',x',0,0)\nonumber\\
&& \quad \quad \approx {4\epsilon\over\pi^2 }\int_{1/\epsilon}^\infty  d x 
\Bigg[-\frac{1}{2(1+x^2)^2}-\frac{ x^2}{4(1+x^2)^2}\nonumber\\
&&\quad \quad\quad\quad\quad\quad\quad\quad
+\frac{8+
8x^2+3x^4}{24(x^2+1)^{5/2}}
\ln\left({\sqrt{x^2+1}+ 1}\over{\sqrt{x^2+1}- 1} \right)\Bigg]\,.\nonumber\\ \,
\een

If we perform the integral and series expand the result, we have, to the order of $O(\epsilon)$,
 
\ben
  &&\int_{a}^{b}dx \int_{a}^b dx'\, G_{xxxx}^{(1) R}(t,x,0,0,\,t',x',0,0)
\nonumber\\
&& \quad \quad \approx c_1^2\epsilon -c_2^2\ln(\epsilon)\,,\nonumber\\
\een
where $c_1$ and $c_2$ are constants given, respectively, by
\be
c_1^2=\int_0^\infty\,dx \frac{\ln(x+\sqrt{x^2+1})}{x\sqrt{x^2+1}}\approx 2.468\,. 
\ee
and 
\be
c_2^2={8\over 3\pi^2}\,.
\ee
Therefore we have for the mean squared geodesic interval fluctuation
\be
\langle \sigma_1^2 \rangle \approx {1\over 8}\left({c_1r\over \pi  }\right)^2\epsilon\,,
\ee
and the mean deviation from the classical propagation time is
\be
\Delta t={\sqrt{\langle \sigma_1^2 \rangle }\over r}\approx {c_1\over \pi}\,{1\over 2\sqrt{2}}
\sqrt{{r\over  L }}.
\ee
This result applies in the regime where $r\gg L$.  Here we have the same functional dependence on $r$ as in
  the case  where light rays loop around the compactified dimension many times.  The only difference lies in the proportionality constants. In fact, here the mean time deviation is also
 smaller than that for light rays traveling in the direction of compactification, 
since the numerical constant $ c_1/\pi\approx 0.5$.


\subsection{Single plane boundary}


Let us assume that there is a single plane boundary located at $z=0$ in space such that metric perturbations satisfy
the following Neumann boundary condition (The reason that we use the Neumann boundary condition instead of the
Dirichlet boundary condition here is to get a positive $\langle \sigma_1^2 \rangle$. )
\be
\partial_z h_{jk}|_{z=0}=0\,.
\ee
In the presence of the boundary, the field mode no longer has the form of 
Eq.~(\ref{eq:mode1} )
but becomes
\be
f_{\bf k} = (\omega (2\pi)^2\pi)^{-{1\over 2}}  e^{i({\bf k_t \cdot x_t} -\omega t)}\cos(k_zz),
	\label{eq: mode2}
\ee
where ${\bf k_t}$ and ${\bf x_t}$ denote the components of ${\bf k} $ and ${\bf x}$, respectively, in
directions parallel to the boundary.
 Now if we assume that the gravitons are in the vacuum state $|0'\rangle $ associated with 
the modes of Eq.~(\ref{eq: mode2}),
 we have, for a light ray propagating perpendicular
to the boundary from point $(0,0,a)$ to $(0,0,b)$, 

\ben
\langle \sigma_1^2 \rangle &&= {1\over 8}(b-a)^2
\int_{a}^{b} dz \int_{a}^{b} dz'
\:\, \langle 0'| h_{zz}(x) h_{zz}(x')+ h_{zz}(x') h_{zz}(x)|0' \rangle_R \,\nonumber\\
&&=\int_{a}^{b} dz \int_{a}^{b} dz'
\:\, G_{zzzz}^{(1)R}(t,0,0,z,\,t',0,0,z') \,.
\label{eq: interval2}
\een

\vskip.25in
\begin{figure}
\begin{center}
\leavevmode\epsfxsize=1.8in\epsfbox{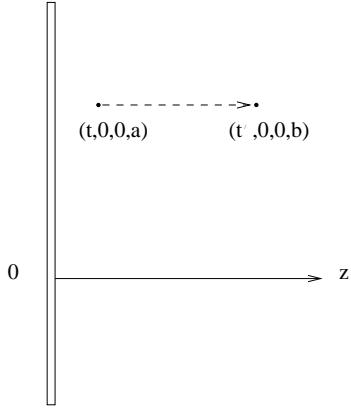}
\end{center}
\caption{ A light ray ( dashed line )propagates in the direction perpendicular to the plane boundary,
starting $a$ distance away from the boundary}
\label{fig=fig4}
\end{figure}
Here the renormalized graviton two point function 
$ G_{zzzz}^{(1)R}(x,x') $
 can be found by the method of images as usual and the only difference is an overall sign change as we go from
  the Dirichlet boundary condition to the Neumann boundary condition. The reason for this is that, to satisfy 
the Neumann boundary condition, we need to add the image term instead of subtracting it as in the case of the 
Dirichlet boundary condition.  So, $ G_{zzzz}^{(1)R}(x,x') $ may be obtained by picking out the 
$n=0$ term in Eq.~(\ref{eq:TPF1}) and setting $\Delta z=z+z'$ to get
\ben
 G_{zzzz}^{(1)R}(t,0,0,z,\,t',0,0,z')&=&-\frac{2(t-t')^2}{\pi^2(z+z')^4}
-\frac{(t-t')^3}{2\pi^2(z+z')^5}\ln\left( \frac{z+z'-(t-t')}{z+z+(t+t')} \right)^2 
\nonumber\\
&&  + {4\over3\pi^2 (z+z')^2}+\frac{(t-t')}{2\pi^2(z+z')^3}\ln\left( \frac{z+z'-(t-t')}{z+z'+(t+t')} \right)^2\,.
\een
Substituting this result into Eq.~(\ref{eq: interval2}) and performing the integration, we finally get
\be
 \langle \sigma_1^2 \rangle =\frac{(b-a)^3\left[ b^2-a^2+(a^2+4ab+b^2)\ln({b\over a})
\right]}
{24\pi^2(b+a)^3}\,.
\ee
Note that this result is always greater than zero. However, had we chosen the Dirichlet boundary condition, we would have
that $\langle \sigma_1^2 \rangle < 0$.  Recall that the
formalism which we are using applies only if $\langle \sigma_1^2\rangle > 0$.

 When the light ray starts very close to the boundary such that $a\ll r$,  
we have
\be
 \langle \sigma_1^2 \rangle\approx {r^4\over 24\pi^2}\left(1+\ln(r/a)\right)\,.
\ee
 The mean deviation in travel time is
\be
\Delta t={\sqrt{\langle \sigma_1^2 \rangle }\over r}=\sqrt{{ 1+\ln(r/a)\over 24\pi^2}}\,,
\label{eq:t1}
\ee
which  diverges as $a$ approaches 0. This is not surprising since the energy density of a 
quantized field blows up on the boundary. However, it has been shown recently \cite{Ford98} that,
 if one treats the boundaries as quantum objects with a nonzero position uncertainty, the 
singularity in energy density is removed.  The result, Eq.~(\ref{eq:t1}), 
  applies whenever $r\gg a$. 
The other 
limit is when $ r\ll a$, where the mean squared fluctuation in the geodesic interval function 
is approximated as 
\be
 \langle \sigma_1^2 \rangle\approx {r^4\over 24\pi^2 a^2}\,,
\ee
 consequently, the mean deviation in time is given by
\be
\Delta t={\sqrt{\langle \sigma_1^2 \rangle }\over r}={1\over 2\sqrt{6}\pi}{r\over a}\,.
\ee

We now consider a null geodesic which is $z$ distance away from and parallel to the plane boundary.
 The relevant renormalized Hadamard function  is
 given by Eq.~(\ref{eq:tpFunc}) with $\Delta z$ being replaced by $z+z'$.
\vskip.25in
\begin{figure}
\begin{center}
\leavevmode\epsfxsize=1.8in\epsfbox{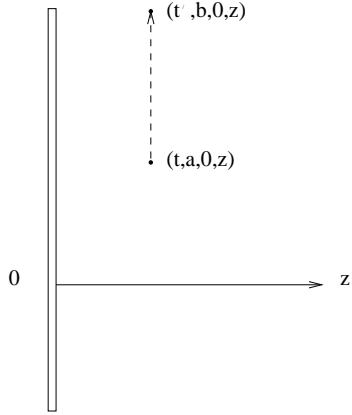}
\end{center}
\caption{ A light ray ( dashed line )propagates in the direction parallel to the plane boundary,
starting $z$ distance away from it }
\label{fig=fig5}
\end{figure}
Now suppose 
the geodesic starts at point $(t,a,0,z)$ and ends at point $(t',b,0,z)$, then the mean squared
 fluctuation in the geodesic interval function is
\be
\langle \sigma_1^2 \rangle = {1\over 8}(b-a)^2
\int_{a}^{b} dz \int_{a}^{b} dz'
\:\, G_{xxxx}^R(t,x,0,z,\,t',x',0,z). \,
\ee
Here
$ G_{xxxx}^R(t,x,0,z,\,t',x',0,z) $ is also given by Eqs.~(\ref{eq:G1}-\ref{eq:G3}) but with a replacement of $nL$ by $2z$.
Therefore
\ben
&&\int_{a}^{b} dx \int_{a}^{b} dx'\, G_{xxxx}^{(1)R }(t,x,0,z,\,t',x',0,z)
\nonumber\\
&&\quad ={2\over\pi^2 }\Bigg[
-\frac{\epsilon^4}{2(\epsilon^2+4)^2}-\frac{\epsilon^2}{(\epsilon^2+4)^2}
\nonumber\\
&&\quad\quad \quad\quad+\frac{8\epsilon^5+
32\epsilon^3+48\epsilon}{24(4+\epsilon^2)^{5/2}}
\ln\left({\sqrt{4+\epsilon^2}+ \epsilon}\over{\sqrt{4+\epsilon^2}- \epsilon} \right)\Bigg]\,,\nonumber\\
\een
where $\epsilon =r/z $. 
Since the above expression is very complicated, we shall discuss two interesting special cases.
 One is when $r\gg z$,  then we have 
 \be
 \langle \sigma^2 \rangle\approx {r^2\over 6\pi^2}\ln(r/z)
\ee
and 
\be
\Delta t={\sqrt{\langle \sigma_1^2 \rangle }\over r}=\sqrt{{\ln(r/ z)\over 6\pi^2}  }\,.
\ee
This also blows up as $z$ approaches 0, however the functional dependence upon $z$ is different from that of
Eq.~(\ref{eq:t1}).
The other limit is when $r\ll z$. For this case, we find
\be
 \langle \sigma^2 \rangle\approx {r^8\over 720z^6\pi^2}
\ee
and 
\be
\Delta t={\sqrt{\langle \sigma_1^2 \rangle }\over r}={1\over 12\sqrt{5}\pi}\, \left({r\over z}\right)^3 \,.
\ee


\section{Summary and Discussion}

  In this paper, we have obtained general expressions, in the transverse tracefree
gauge,  for the vacuum graviton
two-point function for various boundary conditions.
These were used to study the lightcone fluctuations in  flat spacetimes 
with a compactified spatial section and with a plane boundary. The mean squared 
fluctuations of the geodesic interval function and therefore the mean 
deviations from the classical propagation time have been obtained.

 In the case 
of a compactified spatial section, when the travel distance is less than the periodicity 
length, the fluctuation in the propagation time is less than the Planck time. In this limit, 
the effect is much larger for  propagation in the periodicity direction than for propagation 
in the transverse direction. Thus the local lightcone fluctuations become anisotropic, 
reflecting the global structure of the spacetime.   When the travel distance is 
 large  compared to the periodicity length, the fluctuation in travel time increases  with 
the square root of the distance
 traveled for propagation in either direction, and the only difference lies in the 
proportionality constants.   Here we have
 a possibility of having fluctuations larger than Planck scale by several orders of magnitude.

In the case of a  plane boundary,  as light rays start closer and closer to the boundary, 
the lightcone fluctuations blow up as the square root of the 
logarithm  of the starting distance  both when light rays propagate perpendicularly and  parallelly to the boundary.  This 
is not as surprising as it might seem because the imposition of a fixed boundary
can lead to singular expectation values of local observables, such as energy 
densities. However we expect this singularity to disappear if one treats the
boundary as a quantum mechanical object with a nonzero position uncertainty 
\cite{Ford98}. It is also found that if the starting distance from the 
boundary is fixed, then the fluctuation in travel time grows as the square root of the 
logarithm  of the distance traversed when this distance is large compared to the starting 
distance.
 
In summary, we have demonstrated that in the  linearized theory of quantum gravity,
 changes in the topology of flat spacetime produce lightcone fluctuations. These fluctuations 
are in general larger in 
the directions in which topology changes occur and are typically of the order of Planck scale, 
but they can get larger for path lengths large compared to the compactification scale. It is 
interesting to note that this effect  could become significant in theories which postulate
 extra  dimensions compactified on a very small scale.

\begin{acknowledgments}
We would like to thank Tom Roman for interesting discussions and X. Y. Zhong 
for help with graphics. This work was supported in part by the 
National Science Foundation under Grant PHY-9800965.
\end{acknowledgments}

\newpage


\section*{Appendix}
\setcounter{equation}{0}
\renewcommand{\theequation}{A\arabic{equation}}

\subsection{ Summation of graviton polarization tensors in the TT gauge}
Let us introduce a triad of orthonormal vectors $({\bf e }_1({\bf k}) ,
{\bf e }_2({\bf k}) , {\bf e}_3({\bf k})  )$
with
\be
{\bf e}_3({\bf k})  ={{\bf k}\over|{\bf k}|}=\hat{\bf k},
\ee
the unit vector in the direction of propagation. The triad
satisfies the orthonormality relation
\be
{\bf e}_a({\bf k})\cdot{\bf e}_b({\bf k}) =\delta_{ab}, \quad \quad a,b =1,2,3.
\ee
This relation can be written, in terms of the components in the coordinate system
characterizing the metric,  as
\be
e_a^i({\bf k})  e_b^i({\bf k}) =\delta_{ab}, \qquad a,b=1,2,3.
\ee
Here the Einstein summation convention is employed.  We also have
\be
e_a^i({\bf k}) e_a^j({\bf k}) =e^i_1e^j_1+e^i_2e^j_2+\hat k^i\hat k^j
=\delta_{ij}, \qquad  i,j= x,y.z.
\ee
Therefore,  the two independent graviton polarization tensors in the TT gauge
 are given,  in 
 terms of the triad, by 
\ben
&&e^{ij}({\bf k},+)=e^i_1({\bf k}) \otimes e^j_1({\bf k}) 
-e^i_2({\bf k})\otimes e^j_2({\bf k})\,,\\
&&e^{ij}({\bf k},\times)=e^i_1({\bf k})\otimes e^j_2({\bf k})
+e^i_2({\bf k})\otimes e^j_1({\bf k})\,,\\
\een
where we have adopted the notation of Ref \cite{MTW}.
Hence,
\ben
\sum_{\lambda}\, e_{ij} ({{\bf k}, \lambda}) e_{kl} ({{\bf k}, \lambda})&&=
 e_{ij} ({{\bf k}, +}) e_{kl} ({{\bf k}, +})+
 e_{ij} ({{\bf k}, \times}) e_{kl} ({{\bf k}, \times}) \qquad \qquad\nonumber\\
&&=e^i_1e^j_1e^k_1e^l_1-e^i_1e^j_1e^k_2e^l_2-e^i_2e^j_2e^k_1e^l_1+
e^i_2e^j_2e^k_2e^l_2\nonumber\\
&&\quad e^i_1e^j_2e^k_1e^l_2+e^i_1e^j_2e^k_2e^l_1+e^i_2e^j_1e^k_1e^l_2+
e^i_2e^j_1e^k_2e^l_1\nonumber\\
&&=(e^i_1e^k_1+e^i_2e^k_2)(e^j_1e^l_1+e^j_2e^l_2)
+(e^i_1e^l_1+e^i_2e^l_2)(e^j_1e^k_1+e^j_2e^k_2)\nonumber\\
&&\quad -(e^i_1e^j_1+e^i_2e^j_2)(e^k_1e^l_1+e^k_2e^k_2)\nonumber\\
&&=(\delta^{ik}-\hat k^i\hat k^k)(\delta^{jl}-\hat k^j\hat k^l)
+(\delta^{il}-\hat k^i\hat k^l)(\delta^{jk}-\hat k^j\hat k^k)\nonumber\\
&&\quad -(\delta^{ij}-\hat k^i\hat k^j)(\delta^{kl}-\hat k^k\hat k^l)\nonumber\\
&&=\delta_{ik}\delta_{jl}
+\delta_{il}\delta_{jk}-\delta_{ij}\delta_{kl}+\hat k_i\hat k_j \hat k_k\hat k_l  +\hat k_i \hat k_j \delta_{kl}+\nonumber\\
&& \quad \hat k_k \hat k_l \delta_{ij}-\hat k_i \hat k_l \delta_{jk}
-\hat k_i \hat k_k \delta_{jl}-\hat k_j \hat k_l \delta_{ik}-\hat k_j \hat k_k \delta_{il}\,.
	\label{eq:polsum}
\een

This result can also be obtained as follows. 
Let us introduce a 4th-rank tensor
\be
T^{ijkl}({\bf k})=\sum_{\lambda}\, e^{ij} ({{\bf k}, \lambda}) e^{kl} ({{\bf k}, \lambda})\,,
\ee
which has the following symmetry properties
\be
T^{ijkl}=T^{jikl}=T^{ijlk}=T^{klij}.
\ee
However, the objects, which are at our disposal to construct $T^{ijkl}$,
 are only $k^i$ and $\delta^{ij}$, thus in general, we have

\ben
T^{ijkl} &=& A\delta^{ij}\delta^{kl}+B\delta^{ik}\delta^{jl}+B\delta^{il}\delta^{jk}+
C(\hat k^i \hat k^j\delta^{kl}+\hat k^k\hat k^l\delta^{ij})\nonumber\\
&& + D(\hat k^i \hat k^k\delta^{jl}+ \hat k^i \hat k^l\delta^{jk}
+ \hat k^j \hat k^l\delta^{ik}+ \hat k^j \hat k^k\delta^{il})
+E \hat k^i \hat k^j \hat k^k \hat k^l\,,\nonumber\\
\een
where $A,B,C,D,E$ are constants to be determined.
This tensor is subject to the transversality condition
\be
k_iT^{ijkl}=k_jT^{ijkl}=k_kT^{ijkl}=k_lT^{ijkl}=0\,,
\ee
and the trace-free condition
\be
T^{iikl}=T^{ijkk}=0\,.
\ee
Applying these constraint conditions to $T^{ijkl}$ and solving the resulting equations
 leads to 
\be
a=d=-e=-c=-b\,.
\ee
Therefore $T^{ijkl}$ is the same as the right-hand side of Eq.~(\ref{eq:polsum}),  apart
for  a multiplicative normalization constant which can be chosen to be unity.


\subsection{ Vacuum  graviton Hadamard function in the TT gauge}


Here we evaluate the function $F_{ij}(x,x')$ and $H_{ijkl}(x,x')$ defined 
in Eqs.~(\ref{eq:Ffunc}) and (\ref{eq:Hfunc}), respectively.  Once these functions are given, the graviton
two point functions are easy to obtain. 
Define
\be
R=\sqrt{(x-x')^2+(y-y')^2+(z-z')^2},\quad \Delta t=t-t',\quad k=|{\bf k}|=\omega \,.
\ee
Then,
\ben
F_{ij}(x,x')&&={Re\over{(2\pi)^3}}\int\, d^3{\bf k}{k_ik_j\over{2 \omega^3}}e^{i{\bf k} \cdot({\bf x}-{\bf x'})}e^{-i\omega(t-t')} \nonumber\\
&&={Re\over{(2\pi)^3}}\partial_i\partial_j'\int_0^{\infty}\,{e^{-ik\Delta t}\over 2 k}\,dk
\int_0^{\pi}\, d \theta\,\sin\theta e^{i k R \cos\theta}\int_0^{2\pi}\,d\phi\, \nonumber\\
&&={1\over{(2\pi)^2}}\partial_i\partial_j'\,{1\over R} \int_0^{\infty}\,{dk\over k^2} \sin kR\, \cos k\Delta t\,.
\nonumber\\
\een
Because there is an infrared divergence in the above integral, we will introduce a regulator 
$\beta$ in the 
denominator of the integrand and then let $\beta$ approach 0 after the integration is performed.

\ben
F_{ij}(x,x')&&={1\over{(2\pi)^2}}\partial_i\partial_j'\,\lim_{\beta\rightarrow 0}\,{1\over R }  \int_0^{+\infty}\,
{dk\over k^2+\beta^2} \sin kR\, \cos k\Delta t\, \nonumber\\
&&={1\over{(2\pi)^2}}\partial_i\partial_j'\,\lim_{\beta\rightarrow 0}\,  f(\beta, R, \Delta t)\,.
\nonumber\\
\een
Here we have used a integral in Ref.~ \cite{GR1} and  defined
\ben
f(\beta, R, \Delta t)&& = {1\over 4\beta R}\Bigg\{ e^{\beta(\Delta t-R)}{\rm Ei} [\beta(R-\Delta t)]+ e^{-\beta(\Delta t+R)}{\rm Ei} [\beta(R+\Delta t)]\nonumber\\
\quad &&- e^{\beta(\Delta t+R)}{\rm Ei} [-\beta(R+\Delta t)]-e^{\beta(R-\Delta t)}{\rm Ei} [\beta(\Delta t- R)]\Bigg\}\,.\nonumber\\
\een
Here ${\rm Ei(x)}$ is the exponential-integral function.  Making use of the fact that, when $x$ is small, 
\be
{\rm Ei(x)}\approx \gamma +\ln|x|+x+{1\over 4}x^2+{1\over 18}x^3+O(x^4)\,,
\ee
where $\gamma$ is the Euler constant, and expanding $f$ around $\beta=0$ to the order of $\beta^2$, we get
\be
f(\beta, R, \Delta t)\approx 1-\gamma -\ln \beta -{1\over 2}\ln ( R^2-\Delta t^2 )
-{\Delta t \over 4R}\ln \Bigg( \frac{R+\Delta t}{R-\Delta t}\Bigg)^2 +O(\beta^2)\,.
\ee
Taking the limit and keeping in mind that the  constant terms ( with respect to $x$ and $ x'$ ) vanish under 
differentiation, we finally obtain

\be
F_{ij}(x,x')=-{1\over{(2\pi)^2}}\partial_i\partial_j'\,\left[{1\over 2}\ln ( R^2-\Delta t^2 )
+{\Delta t \over 4R}\ln \Bigg( \frac{R+\Delta t}{R-\Delta t}\Bigg)^2  \right]\,.
\ee

Now let us turn our attention to $H_{ijkl}(x,x')$. We have, proceeding with similar steps as we did
for $F_{ij}(x,x')$,

\ben
H_{ijkl}(x,x')&&={Re\over{(2\pi)^3}}\int\, d^3{\bf k}{k_i k_j k_k k_l\over{2 \omega^5}}e^{i{\bf k} \cdot({\bf x}-{\bf x'})}e^{-i\omega\Delta t}\nonumber\\
&&=-{1\over{(2\pi)^2}}\partial_i\partial_j'\partial_k\partial_l'\,
\lim_{\beta\rightarrow 0}\,{1\over 2\beta R }{\partial\over \partial\beta}  \int_0^{+\infty}\,
{dk\over k^2+\beta^2} \sin kR\, \cos k\Delta t\, \nonumber\\
&&=-{1\over{(2\pi)^2}}\partial_i\partial_j'\partial_k\partial_l'\,\lim_{\beta\rightarrow 0}\, {1\over 2 \beta }  {\partial\over \partial\beta}   f(\beta, R, \Delta t)\,.
 \nonumber\\
\een
Now expand ${1\over 2 \beta }  {\partial\over \partial\beta}   f(\beta, R, \Delta t) $ to  order  
$\beta^2$ to find
\ben
{1\over 2 \beta }  {\partial\over \partial\beta}   f(\beta, R, \Delta t)\,&&=-{1\over 2\beta}-{1\over 3}
\left[ (\ln \beta +\gamma-1)R^2+3(\ln \beta +\gamma-1)\Delta t^2\right]\nonumber\\
\quad &&-{1\over 12 R}\left[ (R+\Delta t)^3\ln|R+\Delta t|
+ (R-\Delta t)^3\ln|R-\Delta t|\right]\,.\nonumber\\
\een

Plugging this result into Eq.~(A22) and noting that only terms higher than quadratic in $R$ contribute after
the differentiation, we obtain
\ben
H_{ijkl}(x,x')&=&{1\over 48\pi^2}\partial_i\partial_j'\partial_k\partial_l'\,
\Bigg[ (R^2+3\Delta t^2)\ln (R^2-\Delta t^2)^2\nonumber\\
\quad && + \left(3R\Delta t+{\Delta t^3\over R}\right)\ln \Bigg( \frac{R+\Delta t}{R-\Delta t}\Bigg)^2\Bigg]\,.
\nonumber\\
\een

For convenience, we give the explicit forms for $G_{xxxx}^{(1)}$ and $G_{zzzz}^{(1)}$ here:

\ben
G_{xxxx}^{(1)}(x,x')\,&&=2\left(D^{(1)}(x,x')-2F_{xx}(x,x')+H_{xxxx}(x,x')\right)\nonumber\\
&&={1\over 12\pi^2R^8\sigma^2}\Bigg\{(\Delta x^2-\Delta t^2)(16\Delta x^6-24\Delta x^4\Delta t^2)
-3\Delta t^2 R_t^6\nonumber\\
&&\quad +(9\Delta t^4+69\Delta x^2\Delta t^2+16\Delta x ^4)R_t^4+(-72\Delta x^2\Delta t^4+32\Delta x^4\Delta t^2+32\Delta x^6)
R_t^2\Bigg\} \nonumber\\
&&\quad-{\Delta t\over{16\pi^2R^9}}\ln\left(\frac{R+\Delta t}{R-\Delta t}\right)^2
\Biggl[-R_t^6-(3\Delta t^2 +9\Delta x^2)R_t^4\nonumber\\
&&\quad +24\Delta t^2\Delta x^2R_t^2-8\Delta x^4\Delta t^2
+8\Delta x^6\Biggr]\,,\nonumber\\
	\label{eq:tpFunc}
\een
where 
\ben
&&R_t^2=\Delta y^2+\Delta z^2\\
&&\Delta x =x-x'\\
&&R^2=R_t^2 +\Delta x ^2=\Delta x ^2 +\Delta y^2+\Delta z^2\\
&&\sigma^2=R^2-\Delta t ^2=\Delta x ^2 +\Delta y^2+\Delta z^2-\Delta t ^2\\
\een
To get $G_{zzzz}^{(1)}(x,x')$, all we need to do is to replace $R_t^2 $ in Eq~(\ref{eq:tpFunc}) by 
$R_t^2=\Delta y^2+\Delta x^2$.


\begin{references}

\bibitem{Pauli} W. Pauli, Helv. Phys. Acta. Suppl. {\bf 4}, 69 (1956).
This reference consists of some remarks made by Pauli during the discussion
of a talk by O. Klein at the 1955 conference in Bern, on the 50th anniversary
of relativity theory.

\bibitem{Deser} S. Deser, Rev. Mod Phys. {\bf 29}, 417 (1957).

\bibitem{DW} B. S. De Witt, Phys. Rev. Lett. 13, 114 (1964).
\bibitem{ISS} C.J. Isham, A. Salam, and J. Strathdee, Phys. Rev.  {\bf D3},
1805 (1971); {\bf 5}, 2548 (1972).

\bibitem{MB}  J. D. Bekenstein and V. F. Mukhanov,  Phys. Lett.  {\bf B360}, 7 (1995), gr-qc/9505012.

\bibitem{Ford95} L.H. Ford, Phys. Rev.  {\bf D51}, 1692
               (1995), gr-qc/9410043. 

\bibitem{Ford96} L.H. Ford and N. F. Svaiter, Phys. Rev.  {\bf D54}, 2640
               (1996), gr-qc/9604052

\bibitem{Ford97} L.H. Ford and N. F. Svaiter, Phys. Rev.  {\bf D56}, 2226
               (1997), gr-qc/9704050.

\bibitem{AEMN} G. Amelino-Camelia, J. Ellis,  N.E. Mavromatos and D.V. Nanopoulos, Int. J. Mod. 
Phys.  {\bf A12},  607 (1997). 

\bibitem{EMN} J. Ellis, N.E. Mavromatos and D.V. Nanopoulos, gr-qc/9904068. 

\bibitem{Footnote} Note that this corrects the graviton propagator given in Ref.~\cite{Ford95} ( Eq.~(61) ), 
in which the gauge-dependent term $D_{ij}$ was omitted.

\bibitem{Ford98} L.H. Ford and N. F. Svaiter, Phys. Rev.  {\bf D58}, 065007
               (1998), gr-qc/9804056.

\bibitem{MTW} C.W. Misner, K. Thorne, and J.A. Wheeler, {\it Gravitation},
(W.H. Freeman, San Francisco, 1973), Sect. 35.6.

\bibitem{GR1} I.S. Gradshteyn and I.M. Ryzhik, {\it Table of Integrals,
Series, and Products}, (Academic Press, New York, 1965), p 414, p 737.


\end{references}
\end{document}